\documentclass[aps,twocolumn,superscriptaddress,showpacs]{revtex4}

\usepackage[german,english]{babel}
\usepackage{amssymb}
\usepackage{amsfonts}
\usepackage{amsmath}

\begin{document}

{\bf Comment on ``Anomalous heat conduction and anomalous diffusion in
one-dimensional systems''}

A relation between anomalous diffusion, in which the mean squared 
displacement grows in time like $\langle (\Delta x)^2\rangle=2D_{\alpha}
t^{\alpha}$ ($0<\alpha\le 2$),
and anomalous heat conduction was recently derived through a scaling approach
by Li and Wang \cite{li} (hereafter LW). In this model it is assumed that heat
transport in
a 1D channel is solely due to the flow of non-interacting particles: those
entering the channel from the left have a different average kinetic energy
than those entering from the right. The energies of the particles at both
ends of the channel are defined through the Boltzmann distributions that
correspond to the temperatures of two heat baths coupled to either end. The
authors are correct in stating that different billiard models discussed
in literature belong to this class of processes. However, in this Comment we
point out certain crucial inconsistencies of the LW model with the physical
picture of random processes leading to normal and anomalous diffusion.

Firstly, consider the collision-free heat transport between the two
heat baths. This situation corresponds to ballistic transport, $\alpha=2$,
and the mean first passage time acquires the scaling $T\propto
L/v$. In this case, the model of LW reproduces the original
result \cite{lebowitz} for the heat conductivity, $\kappa\propto L$. The
first inconsistency becomes apparent already in this limiting case: since
the typical velocities of particles entering the channel from the left and
from the right are different, the corresponding left and right mean first
passage time necessarily differ, as well. The equality of both first passage
times invoked in LW can only be fulfilled if the particles are thermalized
{\em within\/} the channel; however, under this assumption, the ballistic
nature is lost and the whole model no longer holds. Partially, this
problem may be circumvented by taking the limiting transition $T_L-T_R\to 0$
in Eq.~(4).

The crucial flaw in LW, as we are going to show now, is the fact that Eq.~(1)
does not necessarily imply Eq.~(2) in the range $0<\alpha\le 2$, and vice
versa. Thus, although it is tempting to argue that if the typical displacement
of the particle grows like $\langle (\Delta x)^2\rangle^{1/2}\propto t^{
\alpha/2}$ then the {\em typical\/}
time for traveling a distance $L$ will scale 
like $\tau\propto L^{2/\alpha}$, one cannot conclude what exactly this time
$\tau$ defines: it may well differ from the mean first passage time, $T$. In
particular, the latter may even diverge while $\tau$ exists.

To explain this need for caution let us first address subdiffusion, which
corresponds to a long-tailed waiting time distribution of the form $\psi
(t)\sim(t/t_0)^{-1-\alpha}/t_0$ ($0<\alpha <1$) \cite{klablushle}. In this
case, it was shown in Ref.~\cite{bvp} that the temporal eigenfunctions for
a finite geometry are given by Mittag-Leffler functions, and therefore the
survival probability decays like $t^{-\alpha}$. Thus, the associated mean
first passage time diverges: $T=\infty$, corresponding to the dominance of
the probability of {\em not\/} moving in subdiffusion \cite{klablushle}.
(We should note that in one of the three references, Ref.~\cite{gittermann},
cited in LW to support their scaling relation, the result for the first
passage time distribution is based on an integral expression, which 
diverges for a waiting time distribution of long-tailed nature, and is
therefore wrong.) Without an external bias, the
conductivity of a subdiffusive system vanishes \cite{bvp,scher}. The other
case, in which the
approach presented in LW fails are L{\'e}vy flights \cite{klablushle}. Their
mean
first passage time exists and is finite, as can be shown using the methods
described in Ref.~\cite{sokolov}; however, their mean squared displacement
{\em diverges\/} \cite{rem}.

It must therefore be concluded that the model proposed in LW is by far less
general than assumed there, and due to the combination of two a priori
unrelated equations contains a crucial flaw in the foundations
such that erroneous results ensue for both subdiffusion and L{\'e}vy flights.
We also note that in a related context a model developed in
Ref.~\cite{denisov} provides analytical and numerical results for the
heat conductivity consistent with our objections.

Finally, we point out that the interpretation in terms of the finite-time
measurement in the case of subdiffusion, brought forth in the Reply of
LW \cite{reply}, would lead to a correct result. However, it would cause
an explicitly cutoff time-dependent mean first passage time, and would
therefore be different from the original model developed in Ref. \cite{li}.

We acknowledge discussions with A. Chechkin, S. Denisov, J. Klafter and
N. Korabel.

\noindent
Ralf Metzler\\
\indent NORDITA -- Nordic Institute for Theoretical Physics\\
\indent Blegdamsvej 17, 2100 Copenhagen {\O}, Denmark

\noindent
Igor M. Sokolov\\
\indent Institut f{\"u}r Physik, Humboldt-Universit{\"a}t zu Berlin\\
\indent Newtonstra{\ss}e 15, 12489 Berlin, Germany


\begin{thebibliography}{99}
\bibitem{rem}
We note that for L{\'e}vy walks, which are often used to describe sub-ballistic
superdiffusion \protect\cite{klablushle},
it is not known what is the exact scaling of the mean first
passage time. Thus, also for superdiffusion it is not clear whether the
result in LW holds.
\end{thebibliography}
\end{document}